\documentclass[aps,pra,superscriptaddress,floatfix,showpacs,10pt, twocolumn,showkeys]{revtex4}
\usepackage{bbm}
\usepackage{amsfonts}
\usepackage[dvips]{graphicx}
\usepackage{amsmath,amsfonts,amssymb,graphics,graphicx,epsfig,color,times,bbm}

\begin{document}

\title{Deterministic generation of large scale atomic W states}
\author{Xue-Ping Zang}
\affiliation{School of Physics {\&} Material Science, Anhui University, Hefei 230601, China}
\affiliation{Department of Mechanical and Electronic Engineering, Chizhou University, Chizhou 247000, China}

\author{Ming Yang\footnote{mingyang@ahu.edu.cn}}
\affiliation{School of Physics {\&} Material Science, Anhui University, Hefei 230601, China}

\author{Fatih Ozaydin\footnote{mansursah@gmail.com}}
\affiliation{Department of Information Technologies, Isik University, Sile, Istanbul, 34980, Turkey}


\author{Wei Song}
\affiliation{Institute for Quantum Control and Quantum Information, School of Electronic and Information Engineering, Hefei Normal University, Hefei 230601, China}

\author{Zhuo-Liang Cao}
\affiliation{School of Physics {\&} Material Science, Anhui University, Hefei 230601, China}
\affiliation{Institute for Quantum Control and Quantum Information, School of Electronic and Information Engineering, Hefei Normal University, Hefei 230601, China}

\begin{abstract}
We present a deterministic scheme for generating large-scale atomic W states in cavity QED system via a simple expansion mechanism, which is realized only by a detuned interaction between two identical atoms and a vacuum cavity mode. With the presented scheme, a W-type Bell pair can be created and an $n$-atom W state can be expanded to a $2n$-atom W state with a unit probability of success in principle. No multi-atom gates, quantum memories or quantum non-demolition measurements are required, greatly simplifying the experimental realization of the scheme. The feasibility analysis shows that our expansion scheme can be implemented with state-of-the-art technologies. Our scheme enables advances not only in quantum information and communication but also in quantum thermodynamics, where atomic W states plays a crucial role.
\end{abstract}



\maketitle

\section{Introduction}
Quantum entanglement not only plays an elementary role in quantum physics, but also has wide applications in quantum information processing (QIP) tasks, for instance quantum key distribution \cite{gisin}, quantum teleportation \cite{bennett1}, quantum metrology \cite{pezze}, quantum cryptography \cite{bennett2} and quantum computation \cite{steane}.
The structure and the characteristics of bipartite entanglement is well understood, while theoretical and experimental exploration of multipartite entanglement is still a great challenge.
Among many interesting properties of multipartite entanglement, a major feature is the presence of inequivalent classes of multipartite entangled states, such as Greenberger-Horne-Zeilinger (GHZ) states \cite{greenberger} and W states \cite{dur1}, and states in one class cannot be converted into a state in another class by local operations and classical communication \cite{dur2}.
The entanglement of W states is more robust against particle losses than that of GHZ states \cite{dur1}.
A W state of appropriate size is a superior candidate for leader election in anonymous quantum networks \cite{hondt} and required for several secure quantum communication protocols \cite{joo1,wang1,cao,liu}.

Quantum coherence of a W state is also unique and introduces new advantages in quantum thermodynamics.
It has recently been found that in the thermalization process of a single mode cavity by three-atom systems, W states achieve the maximum thermalization, in contrast to GHZ states which provide no thermalization \cite{Ceren1}.
This interesting finding suggests that a large-scale atomic W state, as a quantum information molecule, can be good a candidate to be a novel quantum energy resource in quantum heat engines \cite{Ceren2}. The increasing of the size of the quantum system is essential to overcome the decoherence effects and to increase the thermalization capability \cite{Hardal1,Altintas1,Turkpence1}.
Therefore, besides realizing QIP tasks, quantum heat engines require experimentally accessible schemes for preparing large-scale atomic W states.

Several schemes have been presented for preparing and characterizing multipartite entangled W states in various physical systems, for instance cavity quantum electrodynamics (QED) system \cite{deng1}, nuclear magnetic resonance (NMR) system \cite{dogra1,vandersypen,laamme,dogra2}, optical system \cite{eibl1,shi}, superconducting quantum system \cite{deng2,devorer,huang} and ion traps system \cite{duan,roos,haffner}.
However, the achievable sizes of the prepared W states are far from being large-scale.
It was recently found that a W state of polarization encoded photons can be expanded by adding one or two photons at a time via accessing only one photon of the W state \cite{tashima3,tashima4,tashima5,tashima6} and that two EPR pairs \cite{tashima1} and even two arbitrary size W states can be fused to obtain a larger size W state \cite{tashima2} with simple optical setups.
At a cost of integrating three-qubit gates, it is possible to increase the probability of success of the fusion process \cite{Ozaydin3}, or to fuse several W states to arrive at a large-scale W state with a fewer number of fusion processes, to ease the practical implementations \cite{Ozaydin2,Ozaydin1}.
When it comes to atomic W states, we have used light-matter interface to design expansion and fusion schemes for generating large-scale atomic W states in cavity QED \cite{zang1, zang2, zang3}.

Although the fusion and expansion approaches via accessing a single qubit of the state enable preparing large-scale W states, their probabilistic nature still requires at least a sub-exponential resource \emph{cost} with respect to the target size, in terms of the spent entanglement during the preparation process.
What is more, as Ozaydin \emph{et al.} demonstrated via the Monte Carlo simulation in Ref.\cite{Ozaydin1}, the similar scaling of the number of fusion attempts physically limits the experimental accessibility to large-scale W states.
The probabilistic nature of these approaches also requires detection mechanisms for post-selection to find out whether the attempt has been successful or not, not only introducing additional complexity to the system but also shrinking the final size of the prepared W state in the successful attempts.
An additional inherent weakness of most of the above mentioned schemes is that they require prior entanglement, in particular tripartite W states or at least W-type Bell pairs to start and to sustain the process i.e. they actually require another scheme to prepare these primary resources.
Since the \emph{cost} of the generation process in terms of the total entanglement spent throughout the process considerably depends on the \emph{cost} of the primary resources to be spent, Yesilyurt \emph{et al.} proposed a scheme for deterministic generation of W states of four polarization-based entangled photons, to serve as the primary resource \cite{Can2015Acta}.

Very recently,  Yesilyurt \emph{et al.} designed an all-optical expansion mechanism for polarization based entangled photons, which can deterministically expand an $n$-photon W state to a $2n$-photon W state using $n$ ancillary photons \cite{ozaydin4}.
The expansion mechanism used there consists of a controlled-NOT (CNOT) gate and a controlled Hadamard (CH) gate, the latter being decomposable into a CNOT gate and single qubit gates.
Although this mechanism can be realized easily with the current technology for polarization based photonic qubits, this is not the case for the matter qubits, due to the difficulty in experimental realization of the controlled gates.
Therefore, a new, simple, and feasible deterministic expansion mechanism must be designed for preparing matter-qubit W states.

In this paper, we present a simple scheme for deterministically generating large-scale atomic W states by expanding small-size atomic W states in a cavity QED system.
Detuned interactions between a cavity field and two atoms constitute the new deterministic expansion mechanism.
The velocity of the atoms is selected for realizing the deterministic expansion mechanism, which is much easier than realizing the controlled gates, and the feasibility analysis at the end of the paper shows that our scheme can be realized with the current technology.

\section{Expansion mechanism}

The key step of our expanding scheme is the detuned interaction between two identical two-level atoms and a cavity mode.
Assume that two atoms are simultaneously sent through the single-mode cavity, and the interaction Hamiltonian can be described in the interaction picture as \cite{zheng}

\begin{equation}\label{1}
H_i=g\sum_{j=1,2}(e^{-i \delta t} a^\dag S^{-}_j+e^{i\delta t}a S^{+}_j),
\end{equation}
where $S^+_j=|e_j\rangle\langle g_j|$ and $S^-_j=|g_j\rangle\langle e_j|$, with $|g_{j}\rangle$ and $|e_{j}\rangle$ representing the ground and excited states of the $j$th atom, respectively. $a$ and $a^\dagger$ denote, respectively, the annihilation and creation operator of the cavity mode, and $g$ is the coupling strength between the cavity mode and each atom.
The atomic transition frequency is described by the parameter $\omega_0$, $\omega$ is the cavity mode frequency, and $\delta=\omega_0-\omega$ is the detuning between them.
In this scheme, the interaction is detuned, therefor there is no energy exchange between the cavity mode and the atomic system, and the condition is described as $\delta\gg g$.
The effective Hamiltonian can be described as

\begin{equation}\label{2}
H=\lambda[\sum_{j=1,2}(|e_j\rangle\langle e_j|aa^\dag -|g_j\rangle\langle g_j|a^\dag a)+(S^+_1S^-_2+S^-_1S^+_2)],
\end{equation}
where $\lambda=g^2/\delta$. Assume the cavity field is initially prepared in the vacuum state, thus the system effective Hamiltonian reduces to
\begin{equation}\label{3}
H_{eff}=\lambda[\sum_{j=1,2}|e_j\rangle\langle e_j|+(S^+_1S^-_2+S^-_1S^+_2)].
\end{equation}
By solving the Schr\"{o}dinger equation, we can obtain the system evolution of each initial state as \cite{zheng}
\begin{eqnarray}\label{4}
|e_1\rangle|e_2\rangle &\longrightarrow & e^{-i2\lambda t}|e_1\rangle|e_2\rangle,\nonumber \\
|e_1\rangle|g_2\rangle &\longrightarrow & e^{-i\lambda t}(\cos\lambda t|e_1\rangle|g_2\rangle-i\sin\lambda t|g_1\rangle|e_2\rangle), \nonumber \\
|g_1\rangle|e_2\rangle &\longrightarrow & e^{-i\lambda t}(\cos\lambda t|g_1\rangle|e_2\rangle-i\sin\lambda t|e_1\rangle|g_2\rangle), \nonumber \\
|g_1\rangle|g_2\rangle &\longrightarrow & |g_1\rangle|g_2\rangle,
\end{eqnarray}

\noindent constituting the basis of our mechanism.
In the following section, we first present how to use our mechanism to generate a four-atom W state from four atoms initially in a separable state, as an example.

\begin{figure}\label{Fig1}
\centering
\includegraphics[width=0.40\textwidth]{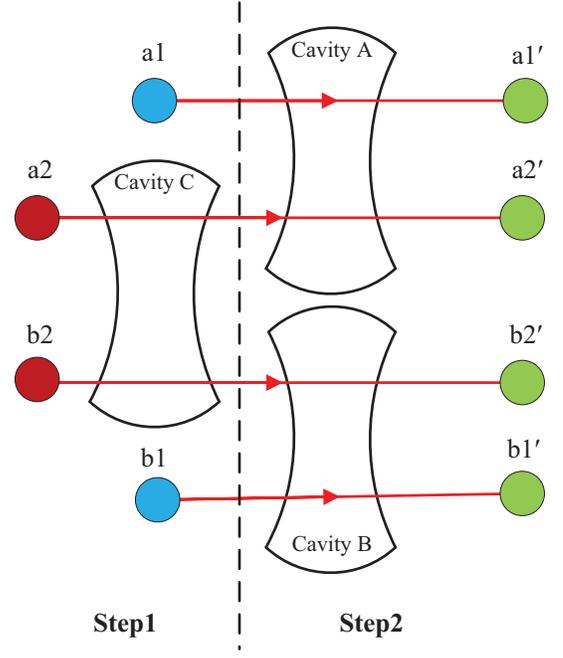}
\caption{(color online). The setup for deterministic generation of a four-atom W state from four atoms illustrated with blue, red, red and blue spheres, respectively, initially in the separable state $|g\rangle_{a1} |e\rangle_{a2} |g\rangle_{b2} |g\rangle_{b1}$.
In Step 1, an EPR pair is prepared, which is then expanded in Step 2 to a W state of four atoms, illustrated with green spheres labeled as $a1'$, $a2'$, $b2'$ and $b1'$.}
\end{figure}

\section{Deterministic generation of four-atom W state via generating and expanding an EPR Pair}

As depicted in FIG.~\ref{1}, there are two steps in our deterministic generation scheme: creating a $W_2$ state, and expanding $W_2$ state to $W_4$ state.
The four two-level atoms ($a1$, $a2$, $b1$ and $b2$) to be entangled are supposed to be identical.
In the first step, atoms $a2$ and $b2$ which are initially in the excited and ground states, respectively, are sent through the cavity $C$ simultaneously.
The detuned interaction between the atoms and the cavity field induces a $W_2$ state, given that the velocity of the atoms are adjusted appropriately.
After these two atoms fly out the cavity $C$, the second step starts, where two ancillary atoms, $a1$ and $b1$ which are initially in ground states, are introduced with adjusted velocities.
Atoms $a1$ and $a2$ are sent through cavity $A$, and atoms $b1$ and $b2$ are sent through cavity $B$, simultaneously.
The expansion process is realized by these two detuned interactions, generating a four-atom W state deterministically.

In more detail, in the first step, the atoms $a2$ and $b2$ are initially prepared in the excited state $|e\rangle_{a2}$ and ground state $|g\rangle_{b2}$, respectively, and their velocities are adjusted so that the detuned interaction time satisfies $\lambda t=\pi/4$.
Via Eq.(\ref{4}), it is easy to see that a $W_2$ state
\begin{equation}\label{5}
|W_{2}\rangle= \frac{1}{\sqrt{2}}(|e\rangle_{a2}|g\rangle_{b2}-i|g\rangle_{a2}|e\rangle_{b2})
\end{equation}
is generated between the two atoms $a2, b2$ as flying out of the cavity $C$, with unit probability.
Here a global phase factor has been omitted.

In the second step, two ancilla atoms $a1, b1$ are both prepared in the ground state ($|g\rangle_{a1}$ and $|g\rangle_{b1}$), and their velocities are adjusted so that the two detuned interactions in cavities $A, B$ have the same duration satisfying $\lambda t=\pi/4$ too.
After leaving the cavities, the state of the four-atom system is found as:

\begin{widetext}
\begin{equation}\label{6}
|W_4\rangle=\frac{1}{\sqrt{4}}(|gegg\rangle_{a1^{'}a2^{'}b1^{'}b2^{'}}-i|egggg\rangle_{a1^{'}a2^{'}b1^{'}b2^{'}}-i|ggge\rangle_{a1^{'}a2^{'}b1^{'}b2^{'}}-|ggeg\rangle_{a1^{'}a2^{'}b1^{'}b2^{'}}),
\end{equation}
\end{widetext}
where the global phase factor has been omitted.
Since the four-atom W state is generated with unit probability, no quantum measurement is required in the generation scheme, therefore the complexity of the scheme is reduced considerably.
A consecutive round of four expansion schemes in parallel can expand this four-atom W state to an eight-atom W state, and so on.
Therefore, in principle, our scheme enables the deterministic generation of any even-number-atom W state, starting with atoms in a separable state, i.e. requiring no prior entanglement.
In the next section, we demonstrate how a $2n$-atom W state can be deterministically expanded from an $n$-atom W state.

\section{Deterministic generation of $2n$-atom W state via expanding an $n$-atom W state}

An $n$-atom $W$ state is expressed as $|W_n\rangle=|(n-1)_g,e\rangle/\sqrt{n}=[|(n-1)_g\rangle_{\widetilde{a1}}\otimes |e\rangle_{a1}+\sqrt{n-1}|W_{n-1}\rangle_{\widetilde{a1}}\otimes |g\rangle_{a1}]/\sqrt{n} $, where the subscript $a1$ denotes the $a1$th atom and $\widetilde{a1}$ denotes the rest $(n-1)$ atoms of $|W_n\rangle$.
The evolution of a $W_n$ state together with $n$ ancilla atoms through $n$ parallel expansion mechanisms can be obtained by the following equations

\begin{equation}\label{7}
|(n-1)_g\rangle_{\widetilde{a1}}|e\rangle_{a1}\otimes|n_g\rangle\longrightarrow|2(n-1)_g\rangle|W_2\rangle,
\end{equation}
\begin{equation}\label{8}
|W_{n-1}\rangle_{\widetilde{a1}}|g\rangle_{a1}\otimes|n_g\rangle\longrightarrow|W_{2(n-1)}\rangle|2_g\rangle.
\end{equation}
Therefore, the final state of the whole system is
\begin{equation}\label{9}
|W_{2n}\rangle=\frac{1}{\sqrt{n}}[|2(n-1)_g\rangle|W_2\rangle+\sqrt{n-1}|W_{2(n-1)}\rangle|2_g\rangle],
\end{equation}
which is a standard $2n$-atom W state.

\section{Results and Discussion}

We have presented a new expansion mechanism which can generate a Bell state from a product state of two atoms, and can double the size of a given atomic W state of two or more atoms, deterministically.
It is well-known that if an entangled W state with odd number of atoms ($|W_{2n+1}\rangle$) is required, one can first deterministically generate the state $|W_{2n+2}\rangle$ and then via a projective measurement on any atom of the state $|W_{2n+2}\rangle$, will be left with the remaining $(2n+1)$ atoms in the state $|W_{2n+1}\rangle$, if the measurement result is $|g\rangle$, with a probability of success $1-1/2(n+1)$ which approximates to $1$ for large $n$.

Since the large detuning interaction between the cavity mode and the atoms adiabatically eliminates the excitation states of cavity mode during the interaction, the cavity decay time does not affect the scheme.
The radiative time of a Rydberg atom with principal quantum number $49$, $50$ or $51$ is $T_r=3\times10^{-2} s$, the atom-cavity coupling strength is $g=2\pi\times24$ kHz \cite{zheng,yang1,brune}, and the choice $\delta=10g$ satisfies the large detuning condition, i.e. the interaction time in each step between the cavity mode and atoms is in the order of $\pi\delta/g^2\simeq2\times10^{-4}s$, implying that the time required for accomplishing the whole process is approximately $10^{-3}s$, which is much shorter than $T_r$.
Therefore, the interaction time of our scheme is much shorter than the radiative time of atoms, and thus the deterministic expansion scheme is realizable with the current cavity-QED technology. Actually, the basic expansion mechanism described in Eq.(\ref{1}), i.e. a coherent control of an atomic collision in a detuned cavity has been experimentally demonstrated \cite{Osnaghi}.
Since the radiative time of the Rydberg atom is about $30$ times that of the interaction time, expansion mechanism can be repeated $30$ times in principle, thus the number of atoms of the final W state can reach up to $2^{30}$.

\section{Conclusion}

In conclusion, a new expansion scheme based on detuned interaction between a vacuum cavity mode and two atoms is proposed to double the size of a given atomic W state, with a unit probability of success.
The scheme can not only expand a given W state but it can also generate a W-type Bell state from a product state of two atoms, allowing one to prepare an arbitrarily large scale W state, starting with atoms initially in a separable state, requiring no other setup for preparing any prior entanglement.
No quantum measurements and no controlled operations are required.
The feasibility analysis shows that our expansion scheme is experimentally realizable with the current cavity-QED technology.
Enabling efficient generation of large-scale atomic W states, our scheme promotes advances in quantum information processing, storage and computation tasks, quantum communication protocols, and the extraction of heat and work in quantum heat engines.
We believe that our work may also open new insights in self-testing of atomic W states.

\begin{acknowledgments}
This work is supported by National Natural Science Foundation of China (NSFC) under Grants No.11274010 and No.11374085, the Key Program of the Outstanding Young Talent of Anhui Province under Grant No.gxyqZD2016370, and the personnel department of Anhui province. F. Ozaydin and M. Yang are funded by Isik University Scientific Research Funding Agency under Grant Number: BAP-15B103.

\end{acknowledgments}

\end{document}